\documentclass[dvips]{ws-p9-75x6-50}

\begin{document}

\title{The current and the charge noise of a single-electron 
transistor in the regime of large charge fluctuations 
out of equilibrium}

\author{Y. Utsumi, H. Imamura, M. Hayashi, and H. Ebisawa}

\address{Graduate School of Information Sciences, Tohoku University, Sendai 980-8579, Japan
\\E-mail:utsumi@cmt.is.tohoku.ac.jp}


\maketitle

\abstracts{
By using the Schwinger-Keldysh approach, 
we evaluate the current noise and the charge noise of 
the single-electron transistor (SET) in the regime
of large charge fluctuations 
caused by large tunneling conductance. 
Our result interpolates between previous theories; 
the \lq \lq orthodox" theory and the \lq \lq co-tunneling theory". 
We find that the life-time broadening effect
suppresses the Fano factor below the value estimated 
by the previous theories. 
We also show that the large tunnel conductance does not reduce
the energy sensitivity so much. 
Our results demonstrate quantitatively 
that SET electrometer can be used as
the high-sensitivity and high-speed device for quantum measurements. 
}

\newcommand{\rd}{{\rm d}}
\newcommand{\ri}{{\rm i}}
\newcommand{\mtau}{\mbox{\boldmath$\tau$}}
\newcommand{\mat}[1]{\mbox{\boldmath$#1$}}

\section{Introduction}
\noindent

A single electron transistor (SET) electrometer is 
an important device for the single shot measurement of a charge qubit
\cite{Devoret} realized 
in the ultrasmall Josephson junction systems\cite{Nakamura1}. 
%
For present day experiment, the dominant mechanism of the decoherence 
of the charge qubit is the $1/f$ back ground charge noise\cite{Nakamura2}, 
which is expected to be reduced in the high-frequency regime 
with technical improvements. 
When the $1/f$-noise is suppressed, the back-action of the measurement, 
i.e. the intrinsic charge fluctuation of SET electrometer, becomes important. 
Especially for a high-speed SET, whose tunneling conductance
is relatively large, 
the charge fluctuation related to the higher-order tunneling process 
is expected to be important. 
So far, the noise has been investigated by using the master equation 
with Markovian approximation, namely \lq \lq orthodox" theory
\cite{KorotkovR,Korotkov1}. 
Beyond the orthodox theory, the quantum fluctuation effect
has been investigated in the Coulomb blockade 
(CB) regime within the second order perturbation theory, 
\lq \lq co-tunneling theory"\cite{Averin,Sukhorukov}. 
Though the semi-quantitative estimation at the threshold has been 
performed\cite{KorotkovR} a decade ago, there is no quantitative 
estimation which covers both the sequential tunneling (ST) 
regime and CB regime. 

Besides the practical application, the noise in
the regime of large quantum fluctuations 
of charge itself is 
interesting from the point of view of the strongly correlated 
system out of equilibrium. 
There have been much development on this topic. 
The renormalization of the conductance and charging energy was predicted
theoretically\cite{Falci,Schoeller_Schon} 
and confirmed experimentally\cite{Joyez}. 
The life-time broadening effect
at finite bias voltage was also predicted\cite{Schoeller_Schon}. 
However how the noise is modified 
by the renormalization effect or the life-time broadening effect
\cite{Schoeller_Schon} has not been clarified. 

In this paper, we derive noise expressions which cover both 
ST regime and CB regime. 
We adopt the modern style of the Keldysh formalism, 
Schwinger-Keldysh approach\cite{Chou,Kamenev_1,Utsumi1}, 
which enables one to calculate any order moment by functional derivative 
of the generating functional. 
We also calculate the energy sensitivity quantitatively 
by using our expressions.

\section{Model and Calculations}
\noindent

\begin{figure}[ht]
\epsfxsize=0.9 \linewidth
\centerline{\epsfbox{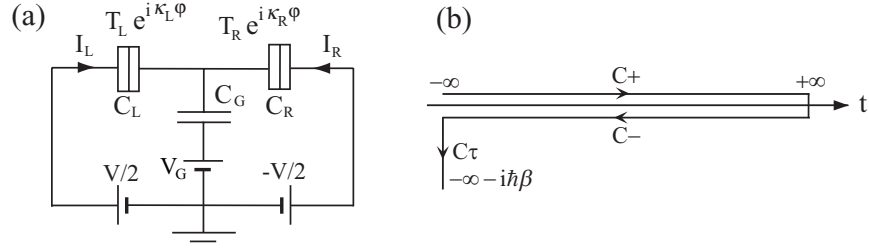}} 
\caption{(a)
The equivalent circuit of a SET transistor. 
(b)
The closed time-path going from $-\infty$ to $\infty$ ($C_{+}$), 
going back to $-\infty$ ($C_{-}$), connecting 
the imaginary time path $C_{\tau}$ and closing at $t=-\infty-\ri \hbar \beta$.
\label{fig:system}}
\end{figure}

Figure \ref{fig:system}(a) shows an equivalent circuit of a SET. 
A normal metal island exchanges electrons with a left 
(right) lead via a small tunnel junction characterized by 
a tunnel matrix element $T_{\rm L(R)}$ and a capacitor $C_{\rm L(R)}$. 
The island is also coupled to a gate via a capacitor $C_{\rm G}$. 
In the following discussion, we limit ourselves to the symmetric case, $C_{\rm L}=C_{\rm R}$ and $T_{\rm L}=T_{\rm R}$. 
We use the two-state model to describe the strong Coulomb interaction. 
%
We begin with the effective action of SET in the drone 
fermion representation\cite{Utsumi1,Isawa}. 
In the limit of large number of transverse channels, 
the effective action for $c$ and $d$-field is given by
\begin{equation}
S
=
\int _C \rd t 
\{
c(t)^* (\ri \hbar \, \partial_t-h(t)) c(t)
+\ri \hbar \, d(t)^* \partial_t d(t)
\}
+
\int _C \rd 1 \rd 2
\{
\sigma_{+}(1)
\alpha(1,2)
\sigma_{-}(2)
\}, 
\label{eq:effective-action}
\end{equation}
where $C$ is the closed time-path (Fig. \ref{fig:system}(b)).
The variable $\sigma_{\pm}$ related to the spin-1/2 operator 
is written by two Grassmann variables 
$c$ and $d$ as $\sigma_{+}=c^* \phi$,
where $\phi=d^*+d$. 
Here $\alpha=\sum_{\rm r=L,R}\alpha_{\rm r}$ and
$\alpha_{\rm r}(1,2)$ is a particle-hole Green function
(GF) which is proportional to the
phase factor ${\rm e}^{\ri \kappa_{\rm r}(\varphi(1)-\varphi(2))}$, where 
$\varphi$ is the phase difference between the left and the right lead. 
$\kappa_{\rm L}=-\kappa_{\rm R}=1/2$ characterizes the voltage drop
between the lead and the island. 
The auxiliary source fields, i.e. the phase difference 
$\varphi$ and the scalar potential for $c$-field $h$ 
are introduced to calculate the average current, the average charge 
and noise by the functional derivative technique. 
It is noticed that the degree of freedom of $\varphi$ ($h$) is duplicated, 
i.e., we can define 
$\varphi_{+}$ ($h_{+}$) and $\varphi_{-}$ ($h_{-}$) 
on the forward $C_{+}$ and the backward branch 
$C_{-}$, respectively. 
The practical form of the particle-hole GF in the 
{\it physical representation}\cite{Chou} is written as
\begin{equation}
\tilde{\alpha}_{\rm r}=\left(
\begin{array}{cc}
0 & \alpha_{\rm r}^A \\
\alpha_{\rm r}^R & \alpha_{\rm r}^K
\end{array}\right),
\; \;
\alpha^R_{\rm r}(\varepsilon)
=
-\ri \pi
\alpha^0_{\rm r}
\rho(\delta \varepsilon^{\rm r}),
\; \;
\alpha^K_{\rm r}(\varepsilon)
=
2 \alpha^R_{\rm r}(\varepsilon)
\coth
\left(
\frac{\delta \varepsilon^{\rm r}}{2T}
\right),
\end{equation}
in the energy space. 
Here 
$\rho(\varepsilon)=\varepsilon$ with Lorentzian cut-off at 
the charging energy $E_C$
and 
$\delta \varepsilon^{\rm r}=\varepsilon-\kappa_{\rm r} eV$. 
The dimensionless junction conductance $\alpha_{\rm r}^0$ 
is defined with the resistance of junction r $R_{\rm r}$ as
$\frac{R_{\rm K}}{(2 \pi)^2 R_{\rm r}}$. 
$\alpha$ is zero on $C_{\tau}$ because 
the tunneling Hamiltonian is adiabatically switched 
on at remote past and off at distant future. 
The generating functional for the connected Green function
$W$ defined as 
$-\ri \hbar \ln \int D \left[c^*,\!c,\!d^*,\!d\right]
\exp \left(\ri S/\hbar \right)$, 
is calculated by performing the perturbation series expansion in powers of 
$\alpha_0$. 
We propose the approximate generating functional including the effect of
infinite order tunneling process
\cite{Utsumi1} 
\begin{equation}
W = - \ri \hbar {\rm Tr}[\ln G_c^{-1}],
\; \; \;
G_c^{-1}(t,t') = g_c^{-1}(t,t')-\sum_{\rm r=L,R} \Sigma_{\rm r}(t,t'),
\label{eqn:W}
\end{equation}
where the trace is performed over $C$. 
Here $g_c^{-1}(t,t')=(\ri \hbar \, \partial_t-h(t)) \, \delta(t,t')$ 
and 
the self-energy is given by 
$\Sigma_{\rm r}(t,t') = -\ri \hbar \, g_{\phi}(t',t) \, \alpha_{\rm r}(t,t')$ 
where 
$g_{\phi}^{-1}(t,t)=\ri \hbar \, \partial_t \delta(t,t')/2$. 
Here, $\delta$-function is defined on $C$ and 
$g_c$ and $g_{\phi}$ satisfy the anti-periodic boundary conditions: 
$g_c(t,-\infty \in C_{+})=-g_c(t,-\ri \hbar \beta-\infty)$,
{\it etc.}

Once an approximate generating functional is obtained, 
one can calculate any order moment by the functional
derivative in terms of auxiliary source fields. 
The averages are calculated as
$I(t)=
\left.
(e/\hbar)
\delta W/\delta \varphi_{\Delta}(t)
\right|_{\varphi_{\Delta},h_{\Delta}=0}$
and 
$Q(t)/e=
1/2-
\left.
\delta W/\delta h_{\Delta}(t)
\right|_{\varphi_{\Delta},h_{\Delta}=0}$
where the relative coordinates, 
$\varphi_\Delta=\varphi_{+}-\varphi_{-}$ {\it etc.}, are set to zero because 
they are fictitious variables. 
The center-of-mass coordinates, $\varphi_c=(\varphi_{+}+\varphi_{-})/2$ 
{\it etc.}, are set as $\varphi_c(t)=eVt/\hbar$ and $h_c(t)=\Delta_0$, 
where $\Delta_0$ is the excitation energy, which is proportional to $E_C$. 
In our previous papers\cite{Utsumi1}, we showed that the resulting 
expressions are formally equivalent to those of 
the resonant tunneling approximation (RTA)\cite{Schoeller_Schon}. 
The noise is defined by the
auto-correlation function of fluctuation operators: 
$
S_{I I}(t,t')=
\langle \{ \delta \hat{I}(t), \delta \hat{I}(t') \} \rangle
$
($\delta \hat{I}=\hat{I}-\langle \hat{I} \rangle$),
{\it etc.} 
In the path integral representation, the noise is calculated by
the second derivative of the generating functional as
\begin{equation}
S_{I I}(t,t')
=
\left.
\frac{1}{\ri \hbar}
\frac{2 \, e^2 \, \delta^2 W}
{\delta \varphi_{\Delta}(t) \delta \varphi_{\Delta}(t')}
\right|_{\varphi_{\Delta},h_{\Delta}=0},
\;
S_{Q Q}(t,t')
=
\left.
\frac{-2 \ri e^2 \hbar \, \delta^2 W}
{\delta h_{\Delta}(t) \delta h_{\Delta}(t')}
\right|_{\varphi_{\Delta},h_{\Delta}=0}.
\end{equation}
In the following discussions, we limit ourselves to the discussions 
on the zero frequency noise, 
$S_{II}=\int \rd t' S_{II}(t,t')$, {\it etc.}
$S_{II}=\sum_{\rm r,r'=L,R} \kappa_{\rm r} \kappa_{\rm r'} 
S_{I{\rm r} \, I{\rm r'}}$, 
is calculated as
$
S_{I{\rm r} \, I{\rm r'}}
=
(e^2/h) {\rm Re}
\int \rd \varepsilon
{\rm Tr}
\left[
\tilde{\Sigma}_{\rm r} \tilde{G}_c 
\, \delta_{\rm r,r'}
+
\tilde{\Sigma}_{\rm r} \mtau^1 \tilde{G}_c \mtau^1 \tilde{\Sigma}_{\rm r'}
\tilde{G}_c
\right.
-
\left.
\tilde{\Sigma}_{\rm r} \mtau^1 \tilde{G}_c
\tilde{\Sigma}_{\rm r'} \mtau^1 \tilde{G}_c
\right]
$, 
where $\mtau^1$ is the Pauli matrix and we omit the argument $\varepsilon$. 
GFs denoted with tilde are those in the physical representation. 
By paying attention to such conditions as 
$\int \rd \varepsilon G_c^{R(A)} \Sigma_{\rm r}^{R(A)}=0$, we obtain the expression for $S_{II} (R_{\rm K}/2)$ written with the Fermi function $f^{-}$ and 
$f^{+}=1-f^{-}$ as
\begin{equation}
\int \rd \varepsilon
[ \,
T^F(\varepsilon)
\{
f^{-}(\delta \varepsilon^{\rm L}) f^{+}(\delta \varepsilon^{\rm R})
+
f^{+}(\delta \varepsilon^{\rm L}) f^{-}(\delta \varepsilon^{\rm R})
\}
-
T^F(\varepsilon)^2
\{
f^{-}(\delta \varepsilon^{\rm L})
-
f^{-}(\delta \varepsilon^{\rm R})
\}^2
\, ], 
\label{eqn:SII}
\end{equation}
%
where the effective transmission probability
$
T^F=-(\alpha_{\rm L}^K \alpha_{\rm R}^K/\alpha^K)
\, 2 \, \ri \, {\rm Im} \, G_c^R
$ 
includes the inelastic scattering process. Here
$G_c^R(\varepsilon)=1/(\varepsilon-\Delta_0-\Sigma_c^R(\varepsilon))$, 
and
$\Sigma_{\rm r}^R(\varepsilon)
=
\alpha_0^{\rm r}
\rho(\varepsilon)
\{
2{\rm Re} \left( \ri \, \frac{\delta \varepsilon^{\rm r}}{2 \pi T} \right)
-
\psi \left( 1+\frac{E_C}{2 \pi T} \right)
-
\psi \left( \frac{E_C}{2 \pi T} \right)
\}
+
\alpha_{\rm r}^K(\varepsilon)/2
$
\cite{Schoeller_Schon}. 
Equation (\ref{eqn:SII}) has the same form as the noise expression of
a point contact without Coulomb interaction\cite{Lesovik}. 
This result is anticipated, because the tunneling current 
is expressed as the Landauer formula with 
the effective transmission probability\cite{Schoeller_Schon}. 
However to our knowledge, there is no literature which derived it
microscopically. 
The charge noise is expressed with off-diagonal components of 
the particle-hole GF 
in the {\it single time representation}\cite{Chou} $\alpha^{\pm \mp}$ as
\begin{equation}
S_{Q Q}
=
e^2 \hbar^2
\int \frac{\rd \varepsilon}{4 h}
{\rm Tr}
\left[
\tilde{G}_c \tilde{G}_c
\right]
=
-\frac{e^4 R_{\rm K}}{2 \pi^2}
\int \rd \varepsilon
\left|
G_c^R(\varepsilon)
\right|^4
\alpha^{-+}(\varepsilon)
\alpha^{+-}(\varepsilon).
\end{equation}

Here we note that our approximation satisfies the minimum required 
properties. 
It is known that the gauge invariance of the generating functional 
leads to the charge conservation\cite{Kamenev_1}. 
In our system, the invariance of 
Eq. (\ref{eqn:W}) under the transformation 
$\varphi_{\rm r} \rightarrow 
\varphi_{\rm r}+\delta \psi$, 
$h \rightarrow
h-\hbar \, \delta (\partial_t \psi)$,
where $\delta \psi$ is defined on $C$, 
leads to the relation, 
$
\partial_t \, \partial_{t'} \, S_{QQ}(t,t')
=
\sum_{\rm r,r'=L,R} S_{I{\rm r} \, I{\rm r'}}(t,t').
$
%
Moreover, one can easily check that Eq. (\ref{eqn:SII}) satisfies
the fluctuation-dissipation theorem at $V=0$. 

\section{Results and Discussions}
\noindent

\begin{figure}[ht]
\epsfxsize=1 \linewidth
\centerline{\epsfbox{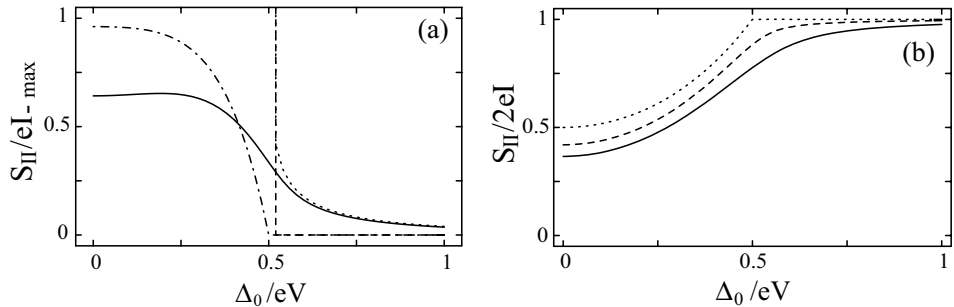}} 
\caption{
(a)
The excitation energy dependence of the current noise 
at 0K and $eV/E_C=0.4$ normalized by $I_{-}=G_0 V/2$, 
where $G_0$ is the series junction conductance. 
The solid and dot-dashed lines show our results Eq. (\ref{eqn:SII}) for 
$\alpha_0=0.05$ and $10^{-5}$, respectively. 
The dotted and dashed lines are results of the co-tunneling theory. 
(b)
The excitation energy dependence of the Fano factor for 
$\alpha_0=0.1$ (solid line), $0.05$ (dashed line) and 
$10^{-5}$ (dotted line). 
\label{fig:1}}
\end{figure}

Figure \ref{fig:1} (a) shows the excitation energy dependence of 
the normalized zero-frequency current noise at finite bias voltage and 
zero temperature. 
When $\alpha_0$ is small, 
our result (the dot-dashed line) reproduce the orthodox theory. 
Moreover, our results are consistent with the 
co-tunneling theory (dotted and dashed lines) 
in the regime $|\Delta_0/eV| \gg 0.5$. 
As $\alpha_0$ becomes large (the solid line), the higher order 
tunneling process enhances 
the current noise around the threshold bias voltage $|\Delta_0/eV|=0.5$. 
However, around $\Delta_0=0$, the current noise is suppressed due to 
the life-time broadening\cite{Schoeller_Schon}. 
The life-time broadening effect is related to the dissipation process 
which is the leak of an electron from the island while another electron 
tunnels into the island.

Figure \ref{fig:1} (b) shows the excitation energy dependence of 
the Fano factor for various $\alpha_0$. 
The Fano factor is the measure how the noise deviates 
from the Poissonian behavior. 
It is known that in ST  
regime $|\Delta_0/eV|<0.5$ the Fano factor is suppressed 
by the Coulomb interaction and that 
in CB regime $|\Delta_0/eV|>0.5$ the Fano factor reaches 1 which means 
that co-tunneling events behave like Poissonian process. 
Our result reproduces this behavior when $\alpha_0$ is small 
(the dotted line). 
As $\alpha_0$ increases, the Fano factor is suppressed over 
ST regime and in CB regime near the threshold 
bias voltage (the dashed line and the solid line). 
In CB regime, the suppression is mainly caused by 
the enhancement of the tunneling probability due to the higher order 
tunneling process. 
On the other hand, around $\Delta_0=0$, the Fano factor suppression 
is caused by the dissipation, i.e. the dissipation regulates tunneling events. 

Above discussions, we consider the condition $eV \gg T_{\rm K}$, 
where $T_{\rm K}$ is the Kondo temperature\cite{Schoeller_Schon}. 
In this regime the renormalization effect is negligible. 
In the opposite regime, the charge noise is suppressed 
due to the renormalization of charge. 

\begin{figure}[ht]
\epsfxsize=1 \linewidth
\centerline{\epsfbox{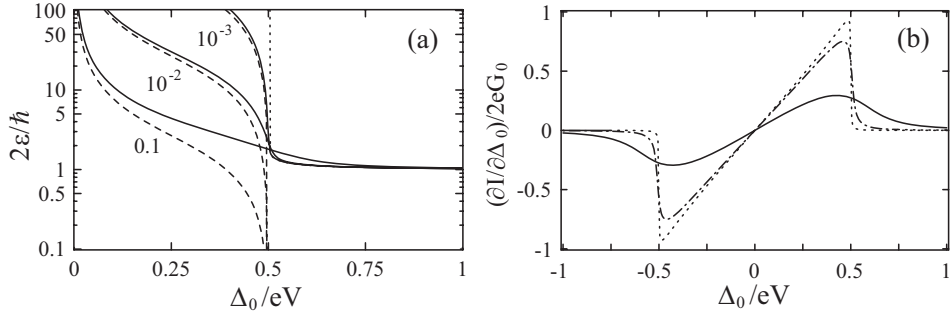}} 
\caption{
(a)
The excitation energy dependence of the energy sensitivity
for $\alpha_0=10^{-3}$, $10^{-2}$ and $0.1$
at 0K and $eV/E_C=0.4$. 
Dashed lines shows results of the orthodox theory. 
The dotted line shows the result of the co-tunneling theory. 
(b)
The slope of the average current. 
The solid, the dot-dashed and the dotted line show the results 
for, $\alpha_0=0.1$, $10^{-2}$ and $10^{-3}$, respectively.
\label{fig:ense}}
\end{figure}

Next we discuss on the performance of SET electrometer. 
The energy sensitivity is defined by the product of the charge noise  
and the charge sensitivity as 
$\epsilon=(\hbar/2) \sqrt{S_{QQ} S_{II}}/|\partial I/\partial \Delta_0|$, 
which is not allowed to be smaller than $\hbar/2$. 
Figure \ref{fig:ense} (a) shows the excitation energy dependence of 
the energy sensitivity at 0 K and $eV/E_C=0.4$ for 
$\alpha_0=10^{-3}$, $10^{-2}$ and $0.1$ (solid lines). 
Three dashed lines show results of the orthodox theory\cite{KorotkovR} 
with corresponding parameters. 
The dotted line shows the result of the co-tunneling theory\cite{Averin} 
which is independent of $\alpha_0$. 
At the threshold, the orthodox theory and 
the co-tunneling theory predict $\epsilon=0$ and
$\epsilon \rightarrow \infty$, respectively. 
Our results interpolate between two theories 
with satisfying $\epsilon>\hbar/2$. 
This fact is considered as an evidence to justify our approximation. 
For the typical value of $E_C=0.1$ meV ($C \sim 800$ af), 
the time constants of SET 
$(R_{\rm T} C)^{-1}=4 \pi \alpha_0 E_C/\hbar$ 
for $\alpha_0=10^{-3}$, $10^{-2}$ and 0.1 
are $1.91$, $19.1$ and $191$ GHz, respectively. 
As seen in Fig. \ref{fig:ense} (a), the energy sensitivity is
at worst $\sim \hbar$ at the threshold which is the usual optimum 
working point. 
Thus our result demonstrate quantitatively 
that the SET electrometer can be operated in the high-frequency regime 
without reduction of the sensitivity
as performed in an experiment\cite{Schoelkopf}. 
Figure \ref{fig:ense} (b) shows the slope of the excitation energy dependence
of the average current for 
$\alpha_0=10^{-3}$ (dotted line), 
$10^{-2}$ (dot-dashed line) 
and $0.1$ (solid line). 
As $\alpha_0$ becomes large, the structure is smeared. 
Our results show 
that the large $\alpha_0$ does not reduce the energy sensitivity 
so much, 
however it makes difficult to obtain the sharp onset of the current.

\section{Summary}
\noindent

%
In conclusion, we have evaluated the noise and the energy sensitivity 
in the regime of 
large quantum fluctuations out of equilibrium. 
%
%
We have reformulated and extended RTA in a charge conserving way. 
Our approximation is justified by the fact that
the current noise satisfies the fluctuation-dissipation theorem 
and 
that the energy sensitivity does not exceed the quantum limit. 
Our approximation consistent with the orthodox theory
in the limit $\alpha_0 \rightarrow 0$, and the co-tunneling 
theory in CB regime. 
%
%
We found that the dissipation, i.e. the life-time broadening effect, 
regulates tunneling events and suppresses the Fano factor. 
%
We have also shown that large $\alpha_0$ does not reduce 
the energy sensitivity so much at the threshold bias voltage. 
Our results demonstrate quantitatively that SET electrometer 
can be used as the high-sensitivity and high-speed device 
for quantum measurements. 



\section*{Acknowledgments}
\noindent

We would like to thank Y. Isawa, J. Martinek and Yu. V. Nazarov 
for variable discussions and comments. 
This work was supported by a Grant-in-Aid for Scientific
Research (C), No. 14540321 from MEXT. H.I. was supported by MEXT,
Grantin-Aid for Encouragement of Young Scientists, No.
13740197.

\end{document}